\documentclass[aps,prep,epsfile,nofootinbib]{revtex4}
\usepackage{graphics}
\usepackage{slashed}
\usepackage{amssymb}
\usepackage{lscape}
\usepackage{amsmath}
\usepackage{graphicx}
\graphicspath{ {images/} }
\begin{document}
\title{Large scales space-time waves from inflation with time dependent cosmological parameter}
\author{$^{2}$ Juan Ignacio Musmarra\footnote{jmusmarra@mdp.edu.ar}, $^{1,2}$ Mauricio Bellini.
\footnote{{\bf Corresponding author}: mbellini@mdp.edu.ar} }
\address{$^1$ Departamento de F\'isica, Facultad de Ciencias Exactas y
Naturales, Universidad Nacional de Mar del Plata, Funes 3350, C.P.
7600, Mar del Plata, Argentina.\\
$^2$ Instituto de Investigaciones F\'{\i}sicas de Mar del Plata (IFIMAR), \\
Consejo Nacional de Investigaciones Cient\'ificas y T\'ecnicas
(CONICET), Mar del Plata, Argentina.}

\begin{abstract}
We study the emission of large-scales wavelength space-time waves during the inflationary expansion of the universe, produced by back-reaction effects. As an example, we study an inflationary model with variable time scale, where the scale factor of the universe grows as a power of time. The coarse-grained field to describe space-time waves is defined by using the Levy distribution, on the wavenumber space. The evolution for the norm of these waves on cosmological scales is calculated, and it is shown that decreases with time.
\end{abstract}
\maketitle

\section{Introduction}\label{1}

The study of the General Relativistic dynamics is a very important topic that has been subject of research since many years ago\cite{Y,GH}. However, during many time, the study of this problem has not evolved too much. The strategy used in \cite{GH}, generalized in \cite{Parattu} for null surfaces, consisted of adding an appropriate term to the original action, so that, when the action was changed, the boundary terms become null. The original problem resides in that, when we consider the Einstein-Hilbert (EH) action ${\cal I}$, which describes gravitation and matter (for $\kappa=8\pi\,G$ --- we shall consider $\hbar=c=1$
throughout the work):
\begin{equation}\label{act}
{\cal I} =\int_V d^4x \,\sqrt{-g} \left[ \frac{R}{2\kappa} + {\cal L}_m\right],
\end{equation}
the variation of the EH action
\begin{equation}\label{delta}
\delta {\cal I} = \int d^4 x \sqrt{-g} \left[ \delta g^{\alpha\beta} \left( G_{\alpha\beta} + \kappa T_{\alpha\beta}\right)
+ g^{\alpha\beta} \delta R_{\alpha\beta} \right]=0,
\end{equation}
includes some boundary terms that cannot be arbitrarily neglected to obtain the Einstein's equations. These terms are the last inside the brackets and must be studied in detail. In some earlier works, we have studied some physical consequences and applications of this terms\cite{rb,rb1,ass,pl1,vts,..}. One of the interesting applications which deserve study is the emission of space-time waves during the inflationary evolution of the universe. Cosmic inflation describes a primordial era in which the universe growth quasi-exponentially and provides a solution to some cosmological problems that cannot be explained otherwise\cite{infl,infl1,infl2,liddle,bcms,mbcs}. During inflation the equation of state $\omega=P/\rho$ remained close to a vacuum expansion: $\omega \simeq -1$, so that the universe became at cosmological scales spatially flat, isotropic, homogeneous, and the energy density was very close to the critical one: $\rho \simeq \frac{3H^2}{8\pi\,G}=\frac{ \lambda}{8\pi\,G}$. This primordial expansion was governed by a unknown kind of energy, named dark energy (a possible explanation about its origin was done in\cite{wh}).

In this work we are aimed to study the space-time waves which came from the local geometrical inhomogeneities during inflation. Back-reaction effects\cite{mau1,mau}, which are described by a massless field $\sigma$, are the geometrical response to the local fluctuations of the inflaton field $\varphi(x)-\left<\varphi(x)\right>=\delta\phi(x)$. In other words, back-reaction effects are the local geometrical space-time distortion due to the local fluctuations of the scalar massive inflaton field. Since these fluctuations are local, on cosmological scales they can be seen as uniformly distributed sources of space-time waves emitted during inflation. This is a very interesting topic to be studied.

The manuscript is organized as follows: In Sect. II we revise and extend Relativistic Quantum Geometry (RQG) by emphasizing the role of the local flux $\delta\Phi$ in the boundary conditions, when we minimize the Einstein-Hilbert action. We obtain the equation of motion for the trace of these space-time waves produced by this flux, and we make a preliminary description of the quantum space-time. In Sect. III we make a description of inflation with a variable time scale. In Sect. IV we study solutions in this inflationary model for back-reaction effects and the trace of space-time waves. In particular, we study solutions on cosmological scales with large wavelengths using a Levy distributions on the momentum space. Finally, in Sect. V we develop some final comments and conclusions.

\section{Relativistic Quantum Geometry with nonzero flux}\label{2}

Now we consider the 4-vector $\delta W^{\alpha}$, given in terms of the varied symmetric connections: $\delta\Gamma^{\alpha}_{\beta\epsilon}$\cite{4}
\begin{equation}\label{w}
\delta W^{\alpha}=
\delta\Gamma^{\epsilon}_{\beta\epsilon}
{g}^{\beta\alpha}-\delta
\Gamma^{\alpha}_{\beta\gamma} {g}^{\beta\gamma}.
\end{equation}
Since we shall consider that the extended manifold given by a displacement of connections, does not preserve the null non metricity on the extended manifold: $\left(g^{\epsilon\nu}\right)_{|\alpha}\neq 0$, will be useful, when we variate an Einstein-Hilbert (EH) action, to consider a flux $\delta \Phi$ of $\delta W^{\alpha}$, as
\begin{equation}\label{fl}
{g}^{\alpha \beta} \delta R_{\alpha \beta}-\delta \Phi =
\left[\delta W^{\alpha}\right]_{|\alpha} - \left(g^{\alpha\beta}\right)_{|\beta}  \,\delta\Gamma^{\epsilon}_{\alpha\epsilon} +
\left(g^{\epsilon\nu}\right)_{|\alpha}  \,\delta\Gamma^{\alpha}_{\epsilon\nu},
\end{equation}
where $"|"$ denotes the covariant derivative on the extended manifold. Here, $\delta R_{\alpha \beta}$ is
\begin{equation}
\delta{R}^{\alpha}_{\beta\gamma\alpha}=\delta{R}_{\beta\gamma}= \left(\delta\Gamma^{\alpha}_{\beta\alpha} \right)_{| \gamma} - \left(\delta\Gamma^{\alpha}_{\beta\gamma} \right)_{| \alpha},
\end{equation}
were we have used an extension of the Palatini identity\cite{pal}.

\subsection{Minimum action with $\delta R_{\alpha\beta}= \lambda(t)\,\delta g_{\alpha\beta}$}

In this work we shall consider the case where $\delta R_{\alpha\beta}$ is related to the variation of the metric tensor
\begin{equation}\label{f2}
\delta R_{\alpha\beta}= \lambda(t)\,\delta g_{\alpha\beta}.
\end{equation}
Here, $\lambda(t)$ is the called cosmological parameter, which is a decaying function of time\cite{dy,mb}. This parameter, takes into account, in an effective manner, the geometrical contribution of back-reaction effects on the background metric. Since these effects have a geometrical origin, they are incorporated in the redefined Einstein tensor: $\bar{G}_{\alpha\beta}= G_{\alpha\beta} - \lambda(t) \,g_{\alpha\beta}$ in such manner that the varied action (\ref{delta}), will can be written as
\begin{equation}\label{del}
\delta {\cal I} = \int d^4 x \sqrt{-g} \left[ \delta g^{\alpha\beta} \left( G_{\alpha\beta} - \lambda(t) \,g_{\alpha\beta} + \kappa T_{\alpha\beta}\right)\right]=0,
\end{equation}
where we have made use of the fact that
\begin{equation}
\delta g^{\alpha\beta}\, g_{\alpha\beta} = - \delta g_{\alpha\beta}\, g^{\alpha\beta}.
\end{equation}
Notice that $\lambda\equiv \lambda(t)$ is not a constant. This is because the extended manifold on which we describe back-reaction effects is not a Riemann manifold and the nonmetricity is not null: $g_{\alpha\beta |\epsilon}\neq 0$. Therefore, vectors and tensors defined on this manifold have a nonconservative norm. Intuitively, we can say that we are dealing with an elastic geometry on this extended manifold. This is necessary to can define gauge invariance on a geometry with "roughness", when we make an effective description on a metric which is isotropic and homogeneous (or, "smooth"). As in previous works\cite{pl1,...}, we shall consider the connections given by Levi-Civita symbols plus a variation that drives the displacement with respect to the Riemann manifold
\begin{equation}\label{ConexionWeyl}
\Gamma^{\alpha}_{\beta\gamma} = \left\{ \begin{array}{cc}  \alpha \, \\ \beta \, \gamma  \end{array} \right\} + \delta \Gamma^{\alpha}_{\beta\gamma} = \left\{ \begin{array}{cc}  \alpha \, \\ \beta \, \gamma  \end{array} \right\}+ b \,\sigma^{\alpha} g_{\beta\gamma}.
\end{equation}
Here, $\sigma_{\alpha}\equiv \sigma_{,\alpha}$ is the ordinary partial derivative of $\sigma$ with respect to $x^{\alpha}$. On the Riemman manifold it is required that non-mectricity to be null: $\Delta g_{\alpha \beta} = {g}_{\alpha \beta ; \gamma} dx^{\gamma}=0$. However, on the extended manifold the variation of the metric tensor is
\begin{equation}
\delta g_{\alpha \beta} = {g}_{\alpha \beta | \gamma} dx^{\gamma} = - b\, (\sigma_{\beta} {g}_{\alpha \gamma} + \sigma_{\alpha} {g}_{\beta \gamma}) dx^{\gamma},
\end{equation}
where ${g}_{\alpha \beta | \gamma}$ is the covariant derivative on the extended manifold given by (\ref{ConexionWeyl}).

For $b=1/3$, it is obtained in (\ref{w}) that $\delta W^{\alpha}=-\sigma^{\alpha}$, and hence the expression (\ref{fl}) can be written as
\begin{equation}\label{gauge}
{g}^{\alpha \beta} \delta R_{\alpha \beta}-\delta \Phi =
\left[\delta W^{\alpha}\right]_{|\alpha} - \left(g^{\alpha\beta}\right)_{|\beta}  \,\delta\Gamma^{\epsilon}_{\alpha\epsilon} +
\left(g^{\epsilon\nu}\right)_{|\alpha}  \,\delta\Gamma^{\alpha}_{\epsilon\nu}= \nabla_{\alpha}\delta W^{\alpha}={g}^{\alpha\beta}\left[ \Box
\delta\Psi_{\alpha\beta}-\lambda(t)\delta g_{\alpha\beta}\right]=-\nabla_{\alpha}\,\sigma^{\alpha}\equiv -\Box \sigma=0,
\end{equation}
where $\delta\Psi_{\beta\gamma}$, can be interpreted as the components of gravitational waves in a more general sense than the standard one. They comply with a tensor-wave equation with a source $\lambda(t)\delta g_{\alpha\beta}$, but its trace is nonzero: $g^{\alpha\beta} \delta\Psi_{\alpha\beta}\neq 0$. Furthermore, in the standard formalism for gravitational waves, the equation of motion is obtained in a linear perturbative expansion with respect to the background and is valid only as a weak field approximation. However, in our work we are dealing with a non-perturbative formalism which is valid for arbitrary gravitational fields. Finally, in the standard formalism the gravitational waves come from a non-conservative quadrupolar momentum, but in our case this is not necessary the case. Notice that eq. (\ref{gauge}) is true only in the case with $b=1/3$, where the flux can be written in terms of a 4-divergence for $\delta W^{\alpha}$ defined in terms of covariant derivatives in the Riemann manifold. To calculate the flux $\delta\Phi$, we must know  $\delta R_{\alpha \beta}$. The variation of the Ricci tensor on the extended manifold is
\begin{equation}\label{VariacionRicciWeyl}
\delta R_{\alpha \beta}  = \left( \delta \Gamma^{\epsilon}_{\alpha \epsilon} \right)_{|\beta} - ( \delta \Gamma^{\epsilon}_{\alpha \beta})_{|\epsilon} \\
 = \frac{1}{3} \left[ \nabla_{\beta} \sigma_{\alpha} + \frac{1}{3} \left( \sigma_{\alpha} \sigma_{\beta} + \sigma_{\beta} \sigma_{\alpha} \right) - {g}_{\alpha \beta} \left( \nabla_{\epsilon} \sigma^{\epsilon} + \frac{2}{3} \sigma_{\nu} \sigma^{\nu} \right) \right],
\end{equation}
so that, in agreement with the equation (\ref{f2}), it is possible to write the left side of (\ref{gauge}) as: $g^{\alpha\beta}\,\left[\delta R_{\alpha\beta} - \lambda(t)\,
\delta g_{\alpha\beta}\right]=0$, and therefore we obtain the most restrictive equation
\begin{equation}\label{ein}
\frac{1}{3} \left[ \nabla_{\beta} \sigma_{\alpha} + \frac{1}{3} \left( \sigma_{\alpha} \sigma_{\beta} + \sigma_{\beta} \sigma_{\alpha} \right) - {g}_{\alpha \beta} \left( \nabla_{\epsilon} \sigma^{\epsilon} + \frac{2}{3} \sigma_{\nu} \sigma^{\nu} \right) \right]-\lambda(t)\,\delta g_{\alpha\beta} =0,
 \end{equation}
where the last term in (\ref{ein}) is due to the flux that cross the closed 3D-hypersurface: $\delta \Phi = \lambda(t)\,g^{\alpha\beta}\,\delta g_{\alpha\beta}$.

In this work we shall consider an expanding universe that is isotropic and homogenous. Due to this fact, it is possible to define the redefined background Einstein equations
\begin{equation}\label{tr}
\bar{G}_{\alpha \beta} = G_{\alpha \beta} - \lambda(t) \,{g}_{\alpha \beta}=-\kappa\, T_{\alpha\beta},
\end{equation}
with the redefined boundary conditions (\ref{ein}). The set of equations, given by (\ref{ein}) and (\ref{tr}), comply with the minimum action principle given in the equation (\ref{del}), because the transformation (\ref{tr}) preserves the EH action, and the flux that cross the 3D-gaussian hypersurface, $\delta\Phi$, is related to the cosmological parameter $\lambda(t)$, and the variation of the scalar field, $\delta\sigma$
\begin{equation}\label{fll}
\delta\Phi = - \frac{2}{3} \lambda(t)\,\delta \sigma.
\end{equation}
The field ${\chi}(x^{\epsilon}) \equiv {g}^{\mu\nu} {\chi}_{\mu\nu}$ is a classical scalar field, such that ${\chi}_{\mu\nu}={\delta {\Psi}_{\mu\nu}\over \delta S\,\,\,\,\,\,\,\,}$ describes the space-time waves produced by the source through the 3D-Gaussian hypersurface \begin{equation}
\Box {\chi} = \frac{\delta \Phi}{\delta S}, \label{bb}
\end{equation}
where $\delta S=U_{\alpha}dx^{\alpha}$. Here, $U^{\alpha}=\frac{dx^{\alpha}}{dS}$ are the components of the 4-velocity, given as a solution of the geodesic equation on the Riemann manifold:
\begin{equation}
\frac{dU^{\alpha}}{dS} + \left\{ \begin{array}{cc}  \alpha \, \\ \beta \, \gamma  \end{array} \right\}\,U^{\beta}U^{\gamma} =0,
\end{equation}
with unity squared norm: $U_{\alpha} \,U^{\alpha}=1$. The differential operator $\Box$ that acts on ${\chi}$ in (\ref{bb}), is written in terms of the covariant derivatives defined on the background metric: $\Box\equiv {g}^{\alpha\beta} \nabla_{\alpha}\nabla_{\beta}$, such that $\nabla_{\alpha}$ give us the covariant derivative on the Riemann background manifold.

\subsection{Quantum space-time}

If we deal with a space-time which is quantum in nature, we can describe it as a Fourier expansion in terms of the modes
\begin{displaymath}
\delta\hat{x}^{\alpha}(t,\vec{x}) = \frac{1}{(2\pi)^{3/2}} \int d^3 k \, \check{e}^{\alpha} \left[ b_k \, \hat{x}_k(t,\vec{x}) + b^{\dagger}_k \, \hat{x}^*_k(t,\vec{x})\right],
\end{displaymath}
where $b^{\dagger}_k$ and $b_k$ are the creation and annihilation operators of space-time, that comply with the algebra $\left< B \left| \left[b_k,b^{\dagger}_{k'}\right]\right| B  \right> = \delta^{(3)}(\vec{k}-\vec{k'})$ and $\check{e}^{\alpha}=\epsilon^{\alpha}_{\,\,\,\,\beta\gamma\delta} \check{e}^{\beta} \check{e}^{\gamma}\check{e}^{\delta}$.
The operators of creation $\delta\hat{x}^{\alpha} (x^{\beta})$, applied to a background state $\left. | B \right>$, return an eigenvalue $dx^{\alpha}$
\begin{equation}
dx^{\alpha} \left. | B \right> =  {U}^{\alpha} dS \left. | B \right>= \delta\hat{x}^{\alpha} (x^{\beta}) \left. | B \right> ,
\end{equation}
so that the background line element results to be
\begin{equation}
dS^2 \, \delta_{BB'}=\left( {U}_{\alpha} {U}^{\alpha}\right) dS^2\, \delta_{BB'} = \left< B \left|  \delta\hat{x}_{\alpha} \delta\hat{x}^{\alpha}\right| B'  \right>,
\end{equation}
and the quantum states are described by a Fock space.

\section{Inflationary universe with time dependent cosmological parameter}

As we have demonstrated in a previous work\cite{vts}, variable time scale can have played an important role in the evolution of the universe.
In order to study an inflationary model where the time scale is variable, we shall consider the line element
\begin{equation}\label{li}
dS^2=e^{-2 \int \Gamma(t) \; dt}dt^2-a_0^2e^{2 \int H(t) \; dt} \delta_{ij} dx^i dx^j.
\end{equation}
Here, the rate of co-moving events is described by the physical time $\tau$. From the point of view of a co-moving relativistic observer, its “clock” evolves as $d\tau=U_0\,dx^0= \sqrt{g_{00}}\,dx^0$. If the expansion of the universe is driven by the inflaton field $\left<\varphi\right>\equiv \phi(t)$ on an isotropic and homogeneous background metric (\ref{li}), the action will be
\begin{equation}
I = \int d^4x \sqrt{-{g}}\left(\frac{{R}}{16 \pi G}-\left[\frac{\dot{\phi}^2}{2}e^{2\int\Gamma(t)dt}-V(\phi)\right]\right).
\end{equation}
The dynamics for the background inflaton field, that drives the expansion of the universe, is
\begin{equation}\label{infdyn}
\ddot{\phi}+\left[3H(t)+\Gamma(t)\right]\dot{\phi}+\frac{\delta \bar{V}}{\delta \phi}=0,
\end{equation}
and the relevant Einstein equations with the time dependent cosmological parameter included, are
\begin{subequations}
\begin{equation}
3H(t)^2+\lambda(t)=8 \pi G \rho,
\end{equation}
\begin{equation}
-[3H(t)^2+2\dot{H}(t)+2\Gamma(t)H(t)+\lambda(t)]=8 \pi G P,
\end{equation}
\end{subequations}
where $P$ and $\rho$ are respectively the pressure and energy density. The equation of state describes the ration between both scalar quantities
\begin{equation}
\omega=\frac{P}{\rho}=-\left(1+\frac{2(\dot{H}(t)+\Gamma(t)H(t))}{3H^2(t)+\lambda(t)}\right),
\end{equation}
such that $\omega$ must remain close to a vacuum dominated state during the inflationary expansion of the universe. Furthermore, from the Einstein equations, we obtain that the redefined scalar potential $\bar{V}(\phi)$ and $\dot{\phi}$, are respectively given by
\begin{subequations}
\begin{equation}
\bar{V}=\frac{e^{-2\int\Gamma(t)}}{8 \pi G}\left(3H(t)^2+\dot{H}(t)+\Gamma(t)H(t)+\lambda(t)\right),
\end{equation}
\begin{equation}\label{phidot}
\dot{\phi}=\sqrt{\frac{-(\dot{H}+\Gamma H)}{4 \pi G}}e^{- \int \Gamma(t) dt}=\sqrt{\frac{p(1-q)}{4 \pi G}}\frac{1}{t^{1+q}}.
\end{equation}
\end{subequations}
Using this in (\ref{infdyn}), we obtain that the following equation must be fulfilled:
\begin{equation}
\dot{\lambda} t^3 - 2 \lambda q t^2 - 12 p^2 q - 2 p q^2 + 2pq = 0.
\end{equation}
The solution for $\lambda(t)$ is
\begin{equation}
\lambda(t) = C_1 t^{2q} + \frac{pq(1-q-6p)}{q+1} \frac{1}{t^2},
\end{equation}
where $C_1$ is a constant. If $q=0$, $\lambda(t)=C_1$ and we can recover a traditional power-law expansion\cite{pl1, kumar}. In this work we shall consider $C_1=0$ and $p=\frac{(1-q)q}{3(3q+1)}$. Then we obtain $\lambda(t)=3H(t)^2=\frac{3p^2}{t^2}$. On the other hand, for $p>0$ we have $q<1$, which is consistent with a real $\dot{\phi}$ in (\ref{phidot}).

\section{Space-time waves from back-reaction effects}

With the choice $b=1/3$, by using the expression (\ref{gauge}), we obtain that
\begin{equation}
g^{\alpha\beta}\,\delta R_{\alpha\beta}=-\left[\Box\sigma+\frac{2}{3} \,\sigma_{\nu}\sigma^{\nu}-\delta\Phi\right]=0,
\end{equation}
and, on the another hand, we have that
\begin{equation}
\nabla_{\alpha} \delta W^{\alpha} \equiv -\Box \sigma=0.
\end{equation}
In order to make resoluble the system of equations, we shall consider the gauge $\sigma_{\nu}\sigma^{\nu}=\lambda(t)\delta\sigma$, where for a co-moving observer $U^0=\sqrt{g^{00}}$, is fulfilled
\begin{equation}\label{esto}
\delta\sigma= U^{\alpha} \sigma_{\alpha}=U^{0} \sigma_{0}.
\end{equation}
With this choice, and using the equations
(\ref{fl}) and (\ref{bb}), the dynamics for $\sigma$ and $\chi$, results to be
\begin{subequations}
\begin{equation}\label{chidynamics}
\Box \sigma = 0,
\end{equation}
\begin{equation}\label{eso}
\Box \chi = - \frac{2}{3}\,U^0 \lambda(t) \dot{\sigma},
\end{equation}
\end{subequations}
where $\Gamma(t)=\frac{q}{t}$ and $H(t)=\frac{p}{t}$. Therefore, in order to solve the dynamics we must first find the solution of (\ref{chidynamics}), to then solve the equation (\ref{eso}), with (\ref{esto}).

In order to describe the fields $\chi$ and $\sigma$, we can expand these fields as Fourier series
\begin{subequations}
\begin{equation}\label{chiexpansion}
\chi(x^\alpha) = \frac{1}{(2\pi)^{\frac{3}{2}}} \int d^3k \; \left[ A_k e^{i\vec{k}.\vec{r}} \Theta_k(t) + c.c.\right],
\end{equation}
\begin{equation}\label{sigmaexpansion}
\sigma(x^\alpha) = \frac{1}{(2\pi)^{\frac{3}{2}}} \int d^3k \; \left[ B_k e^{i\vec{k}.\vec{r}} \xi_k(t) + c.c.\right],
\end{equation}
\end{subequations}
where $\xi_k$ are the time dependent modes of the field $\sigma$, which once normalised, are\cite{vts}:
\begin{equation}\label{xidynamics}
\xi_k(t)=\sqrt{\frac{\pi}{4(p+q-1)}}t^{-\frac{1}{2}(q+3p-1)}{\cal H}^{(2)}_\nu[y(t)].
\end{equation}
Here, ${\cal H}^{(2)}_\nu[y(t)]$ is the second kind Hankel function, with
\begin{subequations}
\begin{equation}
\nu=\frac{q+3p-1}{2(p+q-1)},
\end{equation}
\begin{equation}
y(t)=\frac{k\;t_0^{\;p+q}\;t^{-(p+q-1)}}{a_0\;(p+q-1)}.
\end{equation}
\end{subequations}
The solution can be obtained using the expressions \eqref{chiexpansion}, \eqref{sigmaexpansion} and \eqref{xidynamics} in \eqref{eso}, with the general solution for $\Theta_k(t)$:
\begin{equation}
\Theta_k(t)=\Theta_k^{(h)}(t)+\Theta_k^{(p)}(t).
\end{equation}
Here, the homogeneous part of the solution for the modes of $\chi$, is
\begin{equation}
\Theta_k^{(h)}(t)=C_2\;t^{-\frac{1}{2}(q+3p-1)}\;J_{-\nu}[y(t)]+C_3\;t^{-\frac{1}{2}(q+3p-1)}\;Y_{-\nu}[y(t)].
\end{equation}
The general solution finally results to be
\begin{equation}
\Theta_k^{(p)}(t)=\frac{1}{k^2}\sqrt{\frac{\pi}{9(p+q-1)}}\;\left[h_{1,k}(t)\int\frac{{\cal H}^{(2)}_{\nu_1}[y(t)]}{t^q}\lambda(t)\frac{f_{1,k}(t)}{g_k(t)}dt+h_{2,k}(t)\int\frac{{\cal H}^{(2)}_{\nu_1}[y(t)]}{t^q}\lambda(t)\frac{f_{2,k}(t)}{g_k(t)}dt\right].
\end{equation}
where $J_{-\nu}$ and $Y_{-\nu}$ are the first and second kind Bessel functions with parameter $-\nu$. Furthermore, $h_{1,k}(t)$, $h_{2,k}(t)$, $f_{1,k}(t)$, $f_{2,k}(t)$ and $g_{k}(t)$ are functions given by the expressions
\begin{subequations}
\begin{equation}
h_{1,k}(t)=-t^{-\frac{1}{2}(p-q+1)}\;\frac{a_0}{t_0^{p+q}}\;(p-q+1)\;Y_{\nu_1}[y(t)]-t^{-\frac{1}{2}(q+3p-1)}\;k\;Y_{\nu_2}[y(t)],
\end{equation}
\begin{equation}
f_{1,k}(t)=\frac{a_0}{t_0^{p+q}}\;t^{p+q-1}(p-q+1)\;J_{\nu_1}[y(t)]+k\;J_{\nu_2}[y(t)],
\end{equation}
\begin{equation}
h_{2,k}(t)=t^{-\frac{1}{2}(p-q+1)}\;\frac{a_0}{t_0^{p+q}}\;(p-q+1)\;J_{\nu_1}[y(t)]+t^{-\frac{1}{2}(q+3p-1)}\;k\;J_{\nu_2}[y(t)],
\end{equation}
\begin{equation}
f_{2,k}(t)=\frac{a_0}{t_0^{p+q}}\;t^{p+q-1}(p-q+1)\;Y_{\nu_1}[y(t)]+k\;Y_{\nu_2}[y(t)],
\end{equation}
\begin{equation}
g_k(t)=Y_{\nu_2}[y(t)]J_{\nu_1}[y(t)]-Y_{\nu_1}[y(t)]J_{\nu_2}[y(t)],
\end{equation}
\end{subequations}
with parameters $\nu_1=\frac{q-p-1}{2(p+q-1)}$ and $\nu_2=\frac{p+3q-3}{2(p+q-1)}$, such that they are related by the expression $\nu_2=\nu_1+1$.

In order to obtain a solution to the physical problem in which the waves are produced by the source, the homogeneous solution must be null, so that we impose $C_2=C_3=0$. For an analytical expression of the particular solution, we must approach in the limit case where $y(t) \ll 1$, corresponding to long wavelengths. So we obtain:
\begin{eqnarray}
\Theta_k^{(p)}(t)|_{y(t) \ll 1} \simeq i&&\Gamma(\nu_1)\sqrt{\frac{\pi}{4(p+q-1)}} \nonumber \\
&&\left[(p+q-1)^{\nu_1}\left(\frac{a_0}{k}\right)^{\nu_1-1}\left(\frac{A_1}{t^{3p-2}}-\frac{A_2}{t^{3p+2q}}\right) + (p-q+1)^{\nu_1}\left(\frac{a_0}{k}\right)^{\nu_1+1}\left(\frac{B_1}{t^{p-2q}}-\frac{B_2}{t^{p+2}}\right)\right],
\end{eqnarray}
where the constants $A_1$, $A_2$, $B_1$ and $B_2$ are given by the expressions
\begin{subequations}
\begin{equation}
A_1=C_1(p-q+1)\left(\frac{4t_0^{-(\nu_1-2)(p+q)}}{3\pi(3q+p-3)(5q+3p-5)(p-2q)}+\frac{2^{\nu_1}t_0^{3(p+q)}}{(q+2p-1)[p^2-(q-1)^2]}\right),
\end{equation}
\begin{equation}
A_2=pq(6p+q-1)\left(\frac{4t_0^{-(\nu_1-2)(p+q)}(p-q+1)}{3\pi(3q+p-3)(5q+3p-5)(p+2)}
+\frac{2^{\nu_1}t_0^{3(p+q)}(2p+q-3)}{(3q+2p-1)(q+2p-1)(p+q-1)}\right),
\end{equation}
\begin{equation}
B_1=C_1\left(\frac{4t_0^{-\nu_1(p+q)}(p-q+1)^{\nu_1}(p+q-1)}{3\pi(3q+p-3)(p-2q)(q+1)}+\frac{t_0^{p+q}(3q+p-3)[p^2-(q+1)^2]}{(q+1)[p^2-(q-1)^2]}\right),
\end{equation}
\begin{equation}
B_2=pq(6p+q-1)\left(\frac{4t_0^{-\nu_1(p+q)}(p+q)(p-q+1)^{\nu_1+1}(p+q-1)}{3\pi(3q+p-3)(p+2)}
+\frac{t_0^{p+q}(3q+p-3)[p^2-(q+1)^2]}{(q+1)[p^2-(q-1)^2]}\right).
\end{equation}
\end{subequations}

\subsection{Redefined fields}

The dynamics of $\sigma$ and $\chi$ fields are described respectively by the equations
\begin{subequations}
\begin{equation}
\ddot{\sigma}+[3H(t)+\Gamma(t)]\dot{\sigma}-\frac{e^{-2\int(H(t)+\Gamma(t))dt}}{a_0^2}\nabla^2\sigma=0,\label{si}
\end{equation}
\begin{equation}
\ddot{\chi}+[3H(t)+\Gamma(t)]\dot{\chi}-\frac{e^{-2\int(H(t)+\Gamma(t))dt}}{a_0^2}\nabla^2\chi=-\frac{2\,U^0\lambda(t)\dot{\sigma}}{3\,e^{\int2\Gamma(t)dt}}. \label{ch}
\end{equation}
\end{subequations}
Notice that the right side of the equation (\ref{ch}) is originated in the flux $\delta \Phi$ of $\delta W^{\alpha}\equiv \sigma^{\alpha}$, that cross the 3D-Gaussian hypersurface. In our case, because the relativistic observer is in a co-moving frame the unique nonzero relativistic velocity is $U^{0}$, so that only contributes $\dot\sigma$ in (\ref{ch}). In order to simplify the structure of the equations (\ref{si}) and (\ref{ch}), we make the changes of variables $\sigma=e^{-\frac{1}{2}\int(3H(t)+\Gamma(t))dt}\,u$ and $\chi=e^{-\frac{1}{2}\int(3H(t)+\Gamma(t))dt}\,v$, and we obtain:
\begin{subequations}
\begin{equation}
\ddot{u}+\frac{[\nabla^2-k_0^2(t)]}{a_0^2e^{2\int(H(t)+\Gamma(t))dt}}\,u=0,
\end{equation}
\begin{equation}
\ddot{v}+\frac{[\nabla^2-k_0^2(t)]}{a_0^2e^{2\int(H(t)+\Gamma(t))dt}}\,v=-\frac{2\,U^0\lambda(t)\dot{\sigma}}{3\,e^{-\frac{3}{2}\int(H(t)-\Gamma(t))}}.
\end{equation}
\end{subequations}

\subsection{Coarse-grained}

The coarse-grained approach provides a description of the dynamics for a desirable part of the spectrum. In our case this part is the long-wavelength sector of the spectrum, which is described by wavelengths much bigger than the size of the Hubble horizon. This wavelengths variate with time, because the horizon is expanding. The wavenumber related to the horizon wavelength in a co-moving frame is $k_0(t)$, so that we shall be interested in wavenumbers $k$, which are smaller than $k_0$ in order to describe the cosmological sector (infrared sector), of the spectrum during inflation
\begin{equation}
k \ll k_0(t)\equiv a_0\,e^{\int(H(t)+\Gamma(t))dt}\,\left[\frac{\left[3H(t)+\Gamma(t)\right]^2}{4}+\frac{3\dot{H}(t)+\dot{\Gamma}(t)}{2}\right]^{1/2}.
\end{equation}
At this point, we will define the coarse-grained fields\cite{mbcs}, by using a suppression factor $f(k,t)$ to select the desirable wavenumbers of the infrared
sector of the spectrum:
\begin{subequations}
\begin{equation}
u_{cg} = \frac{1}{(2\pi)^{\frac{3}{2}}} \int d^3k \; f(k,t) \left[ A_k e^{i\vec{k}\vec{r}} \xi_k(t) + c.c. \right],
\end{equation}
\begin{equation}
v_{cg} = \frac{1}{(2\pi)^{\frac{3}{2}}} \int d^3k \; f(k,t) \left[ B_k e^{i\vec{k}\vec{r}} \tilde{\xi}_k(t) + c.c. \right],
\end{equation}
\end{subequations}
where the suppression factor $f(k,t)$ is given by a Levy distribution,
\begin{equation}
f(k,t)=\sqrt{\frac{\epsilon k_0(t)}{2\pi}}\frac{e^{\frac{-\epsilon k_0(t)}{2(k-\epsilon k_0(t))}}}{(k-\epsilon k_0(t))^{\frac{3}{2}}}.
\end{equation}

The square fluctuations for the coarse-grained fields are\cite{liddle}
\begin{subequations}
\begin{equation}
\left< B|u_{cg}^2|B\right>=\int_0^\infty\frac{dk}{k}\mathcal{P}_{u_{cg}}(k)=\frac{1}{2\pi^2}\int_0^{k_0}dk\;k^2\;|\xi_k(t)|^2f(k,t)^2,
\end{equation}
\begin{equation}
\left< B|v_{cg}^2|B\right>=\int_0^\infty\frac{dk}{k}\mathcal{P}_{v_{cg}}(k)=\frac{1}{2\pi^2}\int_0^{k_0}dk\;k^2\;|\tilde{\xi}_k(t)|^2f(k,t)^2.
\end{equation}
\end{subequations}

\begin{figure}
  \centering
    \includegraphics[scale=0.7]{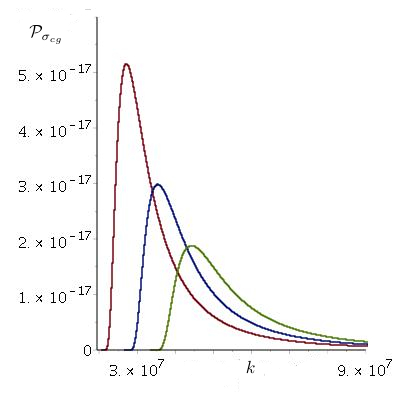}
  \caption{$\mathcal{P}_{u_{cg}}(k)$ for $p=1.5$ and $q=0.9$, with parameters $t_0=G^{1/2}$, $a_0=0.66\,G^{1/2}$ and $\epsilon=10^{-3}$. The red, blue and green lines correspond respectively to the cases $t=1.0\times \;10^{7}\,G^{1/2}$, $t=1.2\times\;10^{7}\,G^{1/2}$ and $t=1.4\times\;10^{7}\,G^{1/2}$.}
  \label{f1gi123}
\end{figure}

\begin{figure}
  \centering
    \includegraphics[scale=0.7]{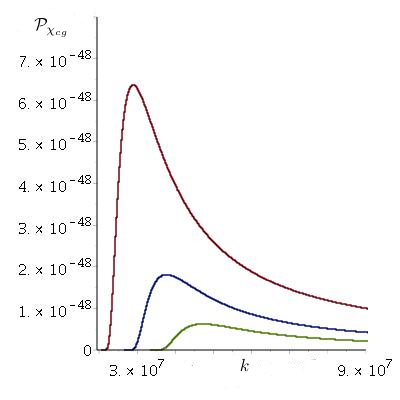}
  \caption{$\mathcal{P}_{v_{cg}}(k)$ for $p=1.5$ and $q=0.9$, with parameters $C_1=0$, $t_0=G^{1/2}$, $a_0=0.66\,G^{1/2}$ and $\epsilon=10^{-3}$. The red, blue and green lines correspond respectively to the cases $t=1.0\times \;10^{7}\,G^{1/2}$, $t=1.2\times\;10^{7}\,G^{1/2}$ and $t=1.4\times\;10^{7}\,G^{1/2}$.}
  \label{f2gi123}
\end{figure}

In the figures (\ref{f1gi123}) and (\ref{f2gi123}) we show the power spectrums of $\mathcal{P}_{\sigma_{cg}}(k)$ and $\mathcal{P}_{\chi_{cg}}(k)$, for different times (during the inflationary era --- time scale is in Planckian times), with $q=0.9$ and $p=1.5$. In both cases the peaks move toward higher $k$-values, while their intensities decrease with time. Notice that the intensities of $\mathcal{P}_{\chi_{cg}}(k)$ are weakest than the $\mathcal{P}_{\sigma_{cg}}(k)$ ones.

\section{Final comments}\label{5}

We have shown that back-reaction effects in the primordial universe act as sources of space-time waves that propagates in all directions. These sources would be homogeneously and isotropically distributed in cosmological scales, which is the scale that concern us. They do not be the standard gravitational waves, but rather space-time waves originated by local space-time fluctuations which have a quantum origin. It is expected that these waves would came from all directions, as cosmic background radiation, but its intensity is so low to be detected with the instrumentation available today. We have calculated the spectrums for the squared fluctuations of $\sigma_{cg}$ and $\chi_{cg}$. In both cases, it is shown that the amplitudes are decreasing with time and the distributions are dispersed along the large-scale $k$-spectrum. In this work we have supposed that these sources are scalar fluctuations. All these sources can be viewed on the background metric as a decaying cosmological parameter $\lambda(t)$, due to the fact we have supposed the universe as globally isotropic and homogenous in the distribution of the sources. However, if we loss homogeneity, this parameter would be a function of $r$ and $t$ [{\it i.e.,} a $\lambda(r,t)$]. This topic will be studied in a future work.

\section*{Acknowledgements}

\noindent The authors acknowledge CONICET, Argentina (PIP 11220150100072CO) and UNMdP (EXA852/18) for financial support.

\newpage


\end{document}